\documentclass[aps,nofootinbib,preprint]{revtex4}

\usepackage{graphicx}
\usepackage{graphics}
\usepackage{amsmath}
\usepackage{subfigure}
\usepackage{color}

\newcommand{\vect}[1]{\mbox{\boldmath${#1}$}}
\begin{document}

\begin{flushright}
RESCEU-7/12 \\
DESY~12-044
\end{flushright}

\title{Evading the pulsar constraints on the cosmic string tension 
in supergravity inflation}

\author{Kohei Kamada$^{1}$}
\email[Email: ]{kohei.kamada"at"desy.de}

\author{Yuhei Miyamoto$^{2,3}$}
\email[Email: ]{miyamoto"at"resceu.s.u-tokyo.ac.jp}

\author{Jun'ichi~Yokoyama$^{3,4}$}
\email[Email: ]{yokoyama"at"resceu.s.u-tokyo.ac.jp}

\affiliation{$^{1}$ Deutsches Elektronen-Synchrotron DESY, Notkestra\ss e 85, D-22607 Hamburg, Germany \\
$^{2}$ Department of Physics, Graduate School of Science, \\
The University of Tokyo, Tokyo 113-0033, Japan \\
$^{3}$ Research Center for the Early Universe (RESCEU), \\
Graduate School of Science, The University of Tokyo, Tokyo 113-0033, Japan \\
$^{4}$ Institute for the Physics and Mathematics of the Universe (IPMU), \\
The University of Tokyo, Kashiwa, Chiba, 277-8568, Japan}


\begin{abstract}

The cosmic string is a useful probe of the early Universe 
and may give us a clue to physics at high energy scales 
which particle accelerators cannot reach. 
Although the most promising tool to observe it is the cosmic microwave background (CMB), 
the constraint from gravitational waves is becoming so stringent 
that detecting its signatures in CMB may be impossible.
In this paper, we construct a scenario that contains cosmic strings 
observable in the 
cosmic microwave background while evading 
the constraint imposed by the recent pulsar timing data.
We argue that cosmic strings with relatively large tension are allowed 
by diluting loops contributing to the relevant frequency range of the gravitational wave background. 
We also present a particle physics model to realize such dilution
in the context of chaotic inflation in supergravity, 
where the phase transition occurs during inflation 
due to the time-dependence of the Hubble induced mass.

\end{abstract}

\maketitle

\section{Introduction}

Cosmic strings \cite{stringreview} are line-like topological defects 
which may have been generated in a phase transition in the early Universe 
through the Kibble mechanism \cite{Kibble:1976sj}\footnote{See 
\cite{Yokoyama:1989pa}, however, for the impossibility of thermal phase transitions to produce strings relevant to structure formation
even in the classical big bang cosmology.}.  
Contrary to other topological defects such as monopoles and domain walls, 
they do not dominate the energy density of the Universe, 
since they evolve according to a scaling law \cite{Kibble:1984hp}. 
At first, they were expected to explain the origin of the primordial density fluctuations that 
lead to structure formation \cite{Vilenkin:1981iu}. 
Although  recent observations of the CMB
have ruled out the possibility that cosmic strings alone are responsible for the primordial density fluctuation, 
they may still contribute to the CMB temperature fluctuation  \cite{Bevis:2007gh,Urrestilla:2008jv,Foreman:2011uj,Dunkley:2010ge},
which would be a clue to exploring physics at high energy scales.

Thus far, many observations for detecting cosmic strings through the CMB 
have been done by using the Gott-Kaiser-Stebbins effect \cite{Kaiser:1984iv}. 
The geodesics of photons passing near a moving string are perturbed, 
which affects the temperature fluctuations of the CMB. 
One can thereby constrain  the line energy density, or tension, $\mu$. 
The latest constraint on the string tension is  $G \mu\lesssim 10^{-7}$ \cite{Bevis:2007gh,Dunkley:2010ge}\footnote{
To be precise, 
the constraint on the string tension depends on the model of string evolution. } with 
$G$ being the gravitational constant. 
Future observations such as Planck \cite{PLANCK_HP,PLANCK_coll} and CMBpol \cite{Baumann:2008aq} 
are expected to detect the signatures of strings 
on small scales or in larger multipoles of the power spectrum
with a tension up to $G \mu \sim 10^{-8}$ \cite{Urrestilla:2008jv,Foreman:2011uj}.

There is yet another probe to detect or constrain cosmic strings, the stochastic gravitational wave background 
(GWB) generated by oscillating string loops
 \cite{Vilenkin:1981bx,Damour:2001bk,Siemens:2006vk,Siemens:2006yp,Olmez:2010bi,Binetruy:2012ze}. 
When long cosmic strings in a Hubble volume intersect with each other, 
they reconnect and generate closed string loops. 
As a consequence, 
the distribution of infinite strings obeys the scaling rule \cite{Albrecht:1989mk, Allen:1991bk, BlancoPillado:2011dq}.
At the same time, the loops produced in such a way oscillate, emit gravitational waves (GWs), and shrink gradually. 
Since the frequency of GWs is determined by the size of cosmic string loops that are generated 
continuously in the cosmic history, the spectrum of GWB, which are the sum of GWs emitted by each loop, 
are expected to range widely, from $10^{-10}$ Hz to $10^5$ Hz or 
even higher frequencies \cite{Vilenkin:1981bx}. 
Although their amplitude is typically very small, we may observe their signatures in the GWB through 
the future gravitational detectors \cite{Binetruy:2012ze} such as eLISA \cite{AmaroSeoane:2012km}, 
DECIGO \cite{DECIGO}, and BBO \cite{BBO} 
as well as the ongoing pulsar timing arrays (PTAs) \cite{Detweiler:1979wn,vanHaasteren:2011ni}. 
In PTAs, GWs can be detected through the modulation of arrival 
of pulses from the millisecond pulsars (MSPs) 
due to the change of the distance between the Earth and MSPs caused by the GWB. 

While the PTAs have not detected the signatures of strings thus far, 
they have obtained stringent constraints on the cosmic string tension \cite{Detweiler:1979wn,vanHaasteren:2011ni}. 
In particular, the European Pulsar Timing Array (EPTA) has recently reported the most stringent constraint, 
$G \mu < 10^{-9}$ \cite{vanHaasteren:2011ni}, though it depends on some assumptions 
regarding string loop formation and 
GW emission\footnote{
Recently, Ref. [28] claims a mild and conservative constraint, 
$G \mu <5 \times 10^{-7}$, for
a certain size of the cosmic string loops, which is disfavored by
recent numerical simulations [22].
For the size of cosmic string loop favored in Ref. [22], Ref. [28]
gives a more stringent
constraint $G \mu < 10^{-10}$.
Although the size of cosmic string loops is still a matter of
debate [29],
we here focus on the case where
the loop length is large as suggested in Ref. [22] and
explore a way
to evade the PTA constraints.
}
Therefore, if we believe this constraint, we cannot hope to detect the signature of cosmic strings in the CMB.

In this paper, we reconsider the formation mechanism of cosmic strings 
and propose a new scenario which does not contradict with the present GW observations 
while accommodating large enough $G\mu$ to be detectable in the CMB. 
If they are sufficiently dilute, the loop formation is reduced and hence 
the amplitude of GWs is suppressed. 
This can be realized if the phase transition takes place during inflation as proposed in several literatures 
\cite{Lazarides:1984pq,Shafi:1984tt,VOS,KL,Yokoyama:1989pa,Yokoyama:1988zza,Nagasawa:1991zr}. 
While previous proposals assumed nontrivial interactions between the string-forming Higgs field 
and the inflaton field \cite{Shafi:1984tt,VOS,KL} or gravity \cite{Yokoyama:1989pa, Yokoyama:1988zza}, 
we here point out that the F-term inflation in supergravity can naturally realize 
a phase transition during inflation thanks to the Hubble induced mass \cite{Dine:1995uk},
as studied in Refs.~\cite{Freese:1995vp,Kamada:2011bt}. 
To be concrete, 
we construct a model that generates cosmic stings during chaotic
inflation \cite{Linde:1983gd}
in supergravity \cite{Kawasaki:2000yn}. 
In this model, the separation of infinite strings is expanded so large that they cannot make loops until
long after inflation and hence they enter the scaling regime at a later epoch, for example, around the 
time of matter-radiation equality. 
Since sources of GWs do not exist at early times, 
we can evade the PTA constraint on the GWB.

The rest of this paper is organized as follows. 
In section \ref{sec2}, 
we briefly review the GWs emitted by cosmic string loops and 
see the current limit from observation using MSPs. 
In section \ref{sec3}, we propose a new scenario for the generation and evolution of cosmic strings 
which is consist with current observations and discuss the possibility of their detection in the future. 
In section \ref{sec4}, we construct a model realizing such a scenario, 
and consider how to determine the parameters of our model. 
Section \ref{sec5} is devoted to discussion.

\section{Gravitational waves from string loops and PTA limit  \label{sec2}}

Here we review the gravitational wave emission from cosmic string loops.
In order to estimate the GWB today, 
first we have to know the size distribution of loops at a given time.
Without any intersection and  reconnection, the energy density 
of long string network would fall 
in proportion to $a^{-2}$.
Actually, however, these processes
 frequently occur on cosmic time scales to transfer
a part of the network's energy to loops which decay by radiating gravitational waves
if the number density of long strings is large enough.
As a result, the system relaxes to a scaling solution and the energy
density of long strings decreases
in proportion to
$d_{\rm H}^{-2}=H^2$.
In terms of the characteristic length of the strings, 
$\xi=\gamma d_{\rm H}=\gamma H^{-1}$, 
where $\gamma$ is a numerical coefficient, we can express the energy density of long strings as
\begin{equation}
\rho=\frac{\xi \mu}{\xi^3}=\frac{\mu}{\xi^2}\,.
\end{equation}
The coefficient $\gamma$ is obtained by numerical simulation.
Recent simulation suggests 
$\gamma=0.15$ in the radiation dominant era,
and $\gamma=0.17$ in the matter dominant era \cite{BlancoPillado:2011dq}.

Now let us determine the loop distribution function $n(t, \ell)$, 
which represents the number density of loops with size $\ell$ at time $t$.
At the moment of loop formation ($t=t_g$), its size is proportional to 
the horizon size. 
We can take the initial length as $\alpha t_g$ with 
$\alpha$ a constant. 
The length of a loop becomes smaller with time due to the loss of 
energy by gravitational wave emission.
The energy transferred from a loop into gravitational waves per unit
time is expressed as $\Gamma G \mu^2$, 
where $\Gamma$ is a constant around 50 \cite{Allen:1991bk}.
Thus, the length of a loop at time $t$ generated at $t_g$ can be written as 
\begin{equation}
\ell =\alpha t_g -\Gamma G \mu(t-t_g) \, .
\end{equation}
In other words, loops with size $\ell$ at time $t$ were formed at
\begin{equation}
t_g =\frac{\ell+\Gamma G \mu t}{\alpha+\Gamma G\mu} \, ,
\end{equation}
while those with size $\ell+d\ell$ at time $t$ were formed at
\begin{equation}
t_g+dt_g=\frac{\ell+d\ell+\Gamma G \mu t}{\alpha+\Gamma G\mu} \, .
\end{equation}
In the flat FRW universe dominated by a perfect fluid with its equation of state given by $P=w\rho$, 
the scale factor evolves as
$a\propto t^{\frac{2}{3(1+w)}}$, and the Hubble parameter is given by $H =\dfrac{2}{3(1+w)t}$.
Taking these facts into account,
we obtain the following functional form.
\begin{eqnarray}
\label{distribution RDMD}
n(t,\ell)
&=&\frac{8(1+3w)}{27(1+w)^3 \gamma^2 t_g^3}\times \frac{1}{\alpha(\ell +\Gamma G \mu t)}\times \left(\frac{a(t_g)}{a(t)}  \right)^3 \notag \\
&\approx&\frac{10(\alpha+\Gamma G \mu)^3}{\alpha(\ell + \Gamma G \mu t)^4}
\times \left(\frac{a(t_g)}{a(t)}  \right)^3\,.
\end{eqnarray}
The last approximation is valid for loops which were generated in 
both  the radiation- ($w=1/3$) and matter-dominated ($w=0$) regimes.

Now we briefly summarize the formulation to estimate the energy density of GWB.
This was originally done by Damour and Vilenkin \cite{Damour:2001bk},
and some improvements were proposed in Ref.~\cite{Olmez:2010bi}, 
whose notation we follow.
Taking the effect of cusps into account, the Fourier mode
of amplitude, $h$, of gravitational radiation emitted
 by a loop with length $\ell$
at redshift $z$ is given by
\begin{equation}
\label{total h}
h(f,z,\ell)= \frac{G \mu H_0\,\ell^{\frac{2}{3}}
}{(1+z)^{\frac{1}{3}}\varphi_r(z)}|f|^{-\frac{4}{3}}
\Theta[1-\theta_m(f,z,\ell)] \,.
\end{equation}
Here $\varphi_r(z)$ represents the comoving distance to the loop defined as
\begin{equation}
\label{r redshift}
    r=\frac{1}{H_0}\int_0^z
    \frac{dz'}{\sqrt{\Omega_{\Lambda} + \Omega_{m,0} {(1+z)}^3+ \Omega_{r,0} {(1+z)}^4 }}
    \equiv \frac{1}{H_0}\varphi_r(z)\,,
\end{equation}
and $\theta_{\rm m}$ is defined by $\theta_{\rm m}\equiv \left(\frac{(1+z)f\ell}{2}\right)^{-\frac{1}{3}}$. 
The argument of the Heaviside function $\Theta$ corresponds to the lowest frequency of loop oscillation.
The energy density of GWB is obtained by integrating the contribution from 
various loops,
\begin{eqnarray}
    \Omega_{\rm GW}(f)&=&\frac{4\pi^2}{3 H_0^2}f^3\int dz\int dl\,
    h^2(f,z,\ell)\frac{d^2R(z,\ell)}{dz d\ell} \notag \\
    &=&\frac{4\pi^2}{3 H_0^2}f^3\int dz\int dh\,
    h^2\frac{d^2R(z,h)}{dz dh}.
\label{omegagw integral}
\end{eqnarray}
Here $\displaystyle{\frac{d^2R(z,\ell)}{dz d\ell}}$ is the observable burst rate per length per redshift 
\begin{equation}
\frac{d^2R(z,\ell)}{dz d\ell}=\frac{1}{H_0^3}\varphi_V(z)\frac{1}{1+z}\frac{2n\left(t(z),\ell\right)}{\ell}
\frac{\theta_m^2(f,z,\ell)}{4} \Theta[1-\theta_m(f,z,\ell)] \, ,
\end{equation}
where $H_0^{-3}\varphi_V(z)dz$ represents the spatial volume corresponding to $z\sim z+dz$.
Its detailed derivation is found in Ref.~\cite{Olmez:2010bi}.

Next, let us consider the observational constraint on the GWB.
Currently the strongest limit comes from the Pulsar Timing Array (PTA) experiment, 
which aims at the detection of the GWB using MSPs. 
(Detailed explanation and current status is found in Ref.~\cite{vanHaasteren:2011ni}.)
This experiment has sensitivity at $f = 10^{-9} \sim 10^{-7}$Hz, 
which is determined by the duration of the observation. 
At this very low frequency, considering the effect of cusps, 
one can conclude that the GWB spectrum of the energy density, 
$\displaystyle{\Omega_{\mathrm{GW}}=\frac{1}{\rho_{cr}}
\frac{d\rho_{\mathrm{GW}}}{d\log f}}$, 
is proportional to $f^{-\frac{1}{3}}$, coming from loops which are generated 
during the matter-dominated era.

According to Ref.~\cite{vanHaasteren:2011ni} the latest limit is 
\begin{equation}
\Omega_{\mathrm{GW}}(f=1\mathrm{yr}^{-1})h^2 \leq 
\begin{cases}
10^{-8.4}(2 \mathrm{\sigma}), \\
10^{-9.0}(1 \mathrm{\sigma}). 
\end{cases}
\end{equation} 
This has been translated to the upper limit on the line density of cosmic string as \cite{vanHaasteren:2011ni},
\begin{equation}
G\mu \leq 4.0 \times 10^{-9}\,.
\end{equation}
Note that this constraint has been obtained using the method of
 Damour and Vikenkin \cite{Damour:2001bk} 
taking $\alpha>\Gamma G\mu$ and incorporating all the relevant 
modes.\footnote{The formalism developed in 
Ref.~\cite{Olmez:2010bi} would yield 
a slightly  more stringent constraint on the line density.}
On the other hand, analysis based on the EPTA data has been done in Ref. \cite{Sanidas:2012ee}
incorporating possible variations of cosmic string properties.
They reported the most conservative constraint on the cosmic string tension as
\begin{equation}
\label{conservativelimit}
G\mu < 5.3\times 10^{-7}\,
\end{equation}
which is obtained in the case of $\alpha$ taking a small value, $\alpha \simeq 10^{-5}$, 
and assuming the effect of cusps is neglegible.
However, recent studies \cite{Damour:2001bk, Siemens:2006vk, Siemens:2006yp,Olmez:2010bi} emphasize that
the effect of cusps, which emit higher mode waves, is unignorable.
Moreover, recent simulations \cite{BlancoPillado:2011dq} indicate that the typical value of $\alpha$ is
much larger than $10^{-5}$. 
If we take the effect of cusps into
account and assume $\alpha\simeq 0.1$ as in
Ref. \cite{Sanidas:2012ee} the constraint of the cosmic string
tension turns out to be much more severe,
$G \mu <10^{-10}$.
Here we propose a method by which the EPTA constraint can be evaded even in
the case where
$\alpha$ is not small and when the effect of cusps are taken into account.

This turns to be possible by considering a different mechanism of string formation.
Since the dominant contribution to $\Omega_{\rm GW}(f=1\mathrm{yr}^{-1})$ in the standard scenario
comes from the loops which were generated during late radiation and early matter-dominated eras, 
we can evade the PTA constraint if the number density of such loops can be reduced. 
In this case, we can explain the PTA constraint without setting $G\mu$ to be small.

\section{A new scenario of string formation  \label{sec3}}

In the standard cosmic string scenario, where strings are formed by spontaneous symmetry breaking after
reheating or  at the end of hybrid inflation \cite{hybrid, Jeannerot:2005mc},
the string network soon relaxes to the scaling solution and it is impossible to reduce the number of small loops
relevant to the PTA constraint.
However, the number of loops can be reduced 
in a scenario where cosmic strings are formed not after inflation but {\it during} inflation. 
This type of scenario was also studied by one of us in a different context \cite{Yokoyama:1989pa}.

Let us see the detail of this scenario.
Since the mean separation of strings becomes larger and larger as inflation proceeds, 
they have to wait for a long time to cross. 
Strings can neither intersect with each other nor form loops until
their mean separation falls well below the horizon distance. 
In this scenario, we can remove the sources of the GWB (cosmic string loops) at early times,
so we can satisfy the observational constraints imposed by GWB with a relatively big value of $G\mu$.

We calculate $\Omega_\mathrm{GW}$ in such a scenario 
where loops start to  form only after the time $t_\mathrm{sc}$, 
when strings begin to satisfy the scaling law. 
Although some loops may have been formed before $t_\mathrm{sc}$, 
such loops would be rare and hence 
we here evaluate the GWB spectrum assuming that they are negligible
and the GW emission starts at $t_{sc}$ suddenly.
 
The calculation can be done as follows.
We perform $z$ integral in Eq.~\eqref{omegagw integral} 
to $z_{\rm sc}$, which corresponds to the redshift of
$t_\mathrm{sc}$.
The relation between cosmic time $t$ and $z$ is given by
\begin{eqnarray}
t&=&\frac{1}{H_0}\int_{z}^{\infty}
\frac{dz'}{(1+z')\sqrt{\Omega_{\Lambda 0}+(1+z')^3\Omega_{\rm m 0}
+(1+z')^4\Omega_{\rm r 0}}}
\equiv \frac{1}{H_0} \varphi_t(z), \\
z&=&\varphi_z(H_0t)\quad (\mathrm{the\; inverse\; function\; of\:}\varphi_t(z)).
\end{eqnarray}
Some modifications are needed in the $h$ integral in Eq.~\eqref{omegagw integral} .
Shortly after $t_\mathrm{sc}$, at time $t_1$,
the minimum loop size is $\ell_1 \equiv \alpha t_{\rm sc}-\Gamma G \mu (t_1-t_\mathrm{sc})$ 
since very small loops are absent. 
Besides, the $\Theta$-function means that only the strings with $\ell > \frac{2}{f(1+z)} \equiv \ell_2$ 
contribute to the GWB.  
Therefore the
$h$ integral runs not from 0 but form max($h_1, h_2$), where $h_1$ and $h_2$ correspond to the amplitude of GWs 
emitted from the strings with the length $\ell_1$ and $\ell_2$, respectively. 
The upper limit of $h$ corresponds to the maximum loop size.
Although the authors of Ref.~\cite{Siemens:2006yp} insist that one can remove 
the contribution from rare bursts by introducing the cutoff in
the $h$ integral, the value of the cutoff is ambiguous.  
Therefore we do not introduce it and include ``all bursts''.

\begin{figure}[htbp]
\begin{center}
\begin{minipage}{0.9\hsize}
  \begin{center}
  \includegraphics[width=130mm]{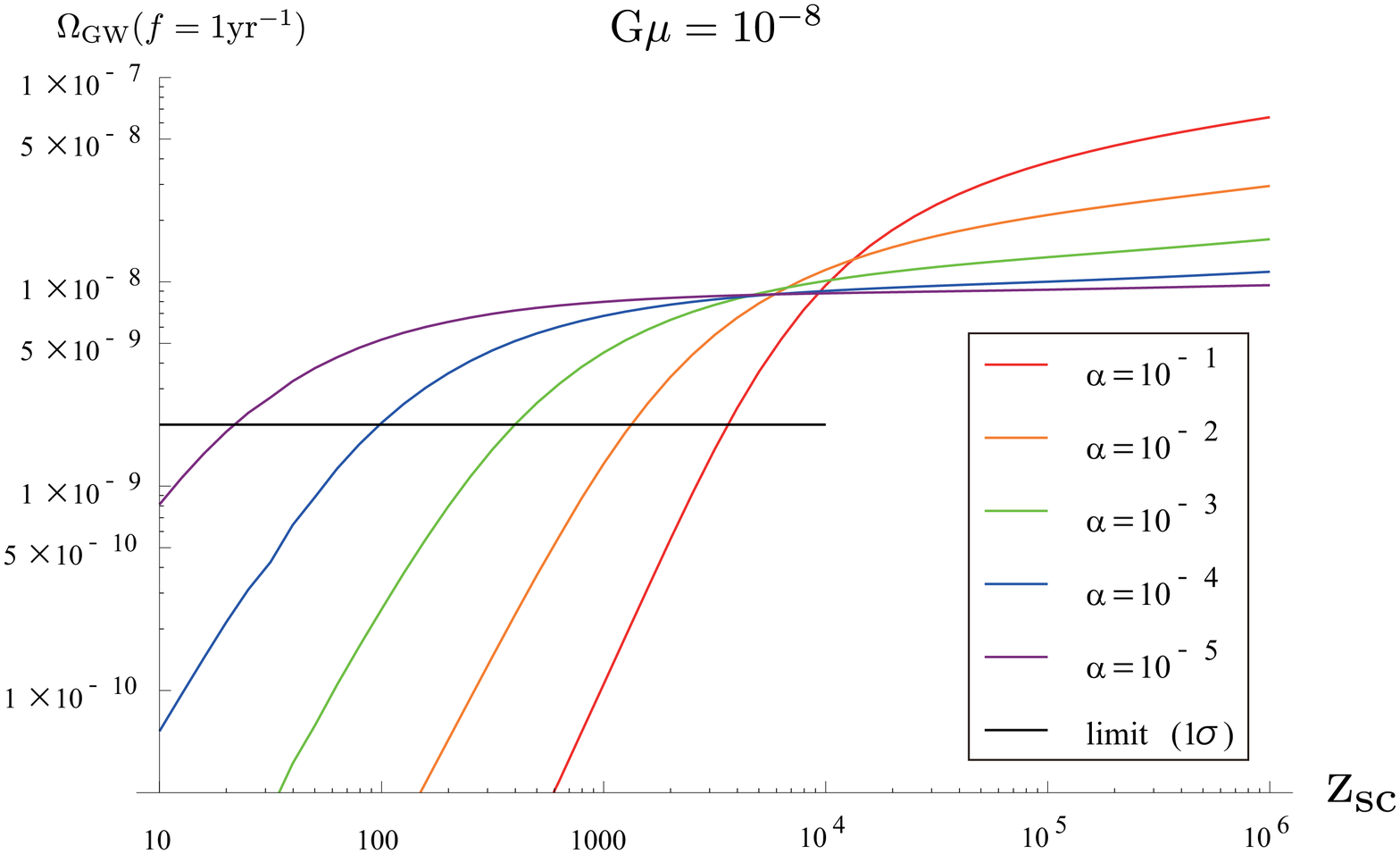}
  \end{center}
\vspace{1cm}
  \begin{center}
  \includegraphics[width=130mm]{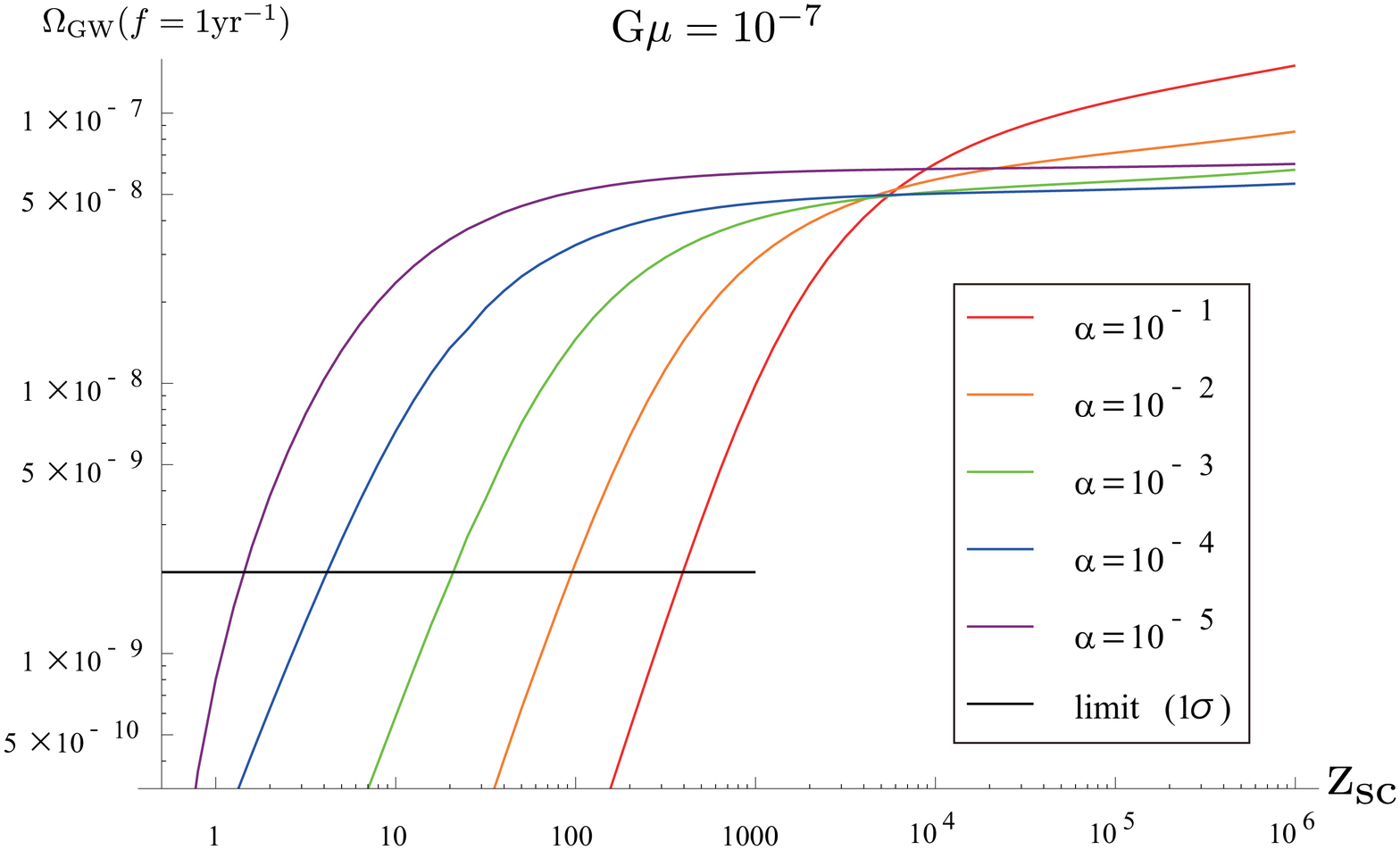}
  \end{center}
 \caption
 {Energy density of GWB at $f={\rm 1year}^{-1}$, $\Omega_{\rm GW}$, 
 as a function of $z_{\rm sc}$, the redshift at the
 onset of the scaling solution
 for various values of $\alpha$. The horizontal line shows the
upper bound imposed by the PTA observation. \label{fig1}}
\end{minipage}
 \end{center}
\end{figure}

Figure \ref{fig1} shows the density parameter of GWBs from cosmic 
strings at the frequency $f=1~{\rm year}^{-1}$
as a function of $z_{\rm sc}$ for 
various values of $\alpha$. 
We can read off the condition
$z_{\rm sc}\simeq 1-10^3$ should satisfy for each value of $\alpha$ and 
$G\mu$ to be consistent with the PTA observation.
Figure 2 shows the maximum $z_{\rm sc}$ 
(or the earliest time for strings to start forming loops) 
in order not to contradict with the PTA data for $\alpha=10^{-1}$.
We find an approximate formula for $z_{\rm sc}$, which evades the PTA limit as 
\begin{equation}
z_{\rm sc} \lesssim 3\times 10^3 \left(\frac{G\mu}{10^{-8}}\right)^{-0.94}\,.
\end{equation}
Therefore, we still have the opportunity to detect features of cosmic string using 
the future CMB observations with small enough $z_{\rm sc}$ since strings with 
$G\mu \lesssim 10^{-7}$ are allowed if loop formation is delayed.

\begin{figure}[htbp]
 \begin{center}
  \includegraphics[width=120mm]{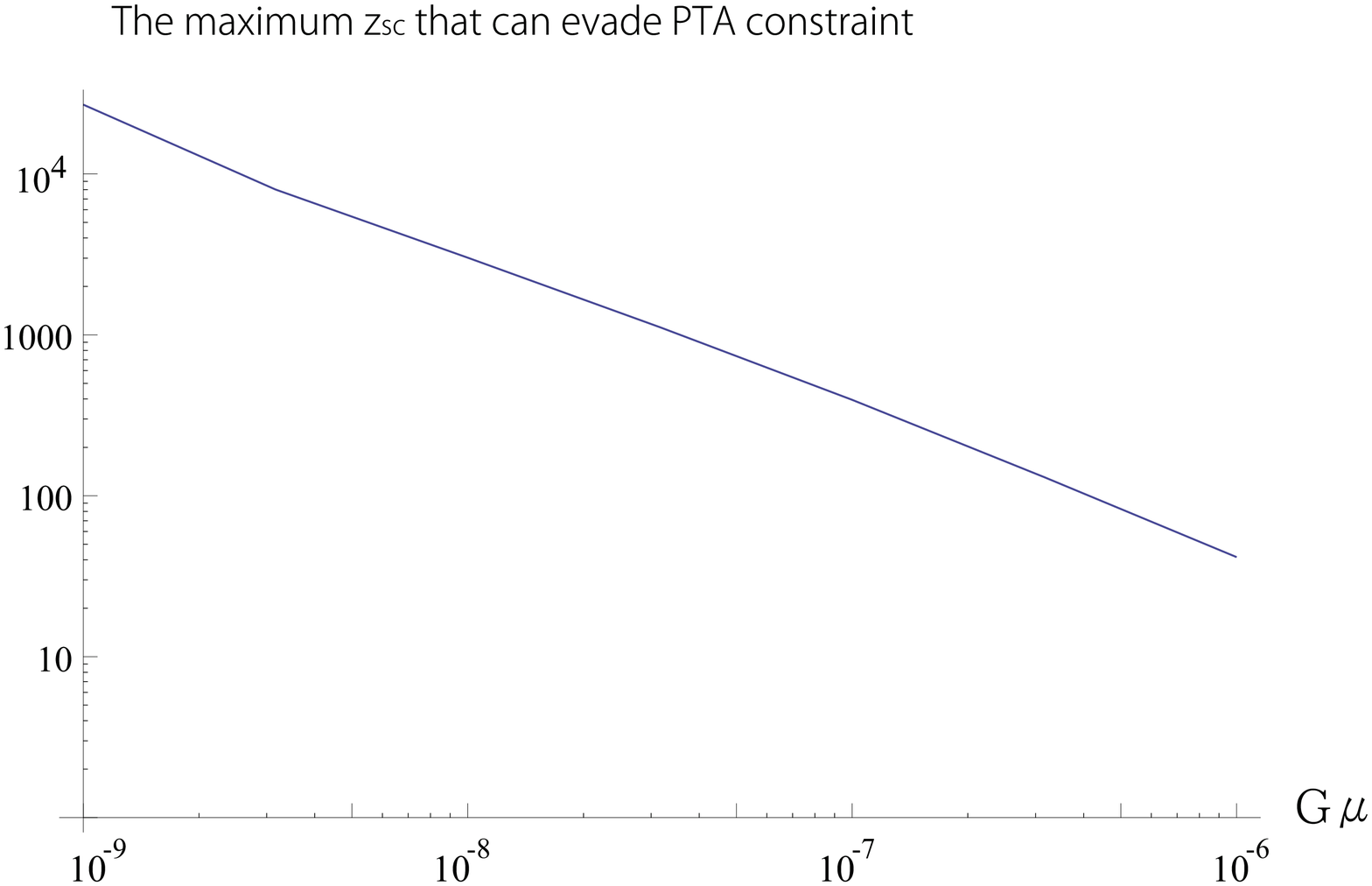}
 \end{center}
 \begin{minipage}{0.9\hsize}
 \caption{The maximum $z_{\rm sc}$ that can evade the PTA constraint as a function of $G\mu$ for $\alpha=10^{-1}$.}
 \end{minipage}
 \label{20120413maxzsc}
\end{figure}

\section{Cosmic strings and inflation in supergravity \label{sec4}}

We present a new mechanism to realize the above scenario in supergravity.
While the previous models introduced a 
nonminimal coupling of the scalar curvature 
to the string-forming scalar field \cite{Yokoyama:1989pa,
Yokoyama:1988zza} or direct coupling with the inflaton
\cite{Shafi:1984tt, VOS,KL}, 
our model does not require any additional interactions, 
since it is automatically provided by gravitationally suppressed interaction arising from supergravity.

The F-term potential in supergravity is written as 
\begin{equation}
V=e^{\frac{K}{{M_{\rm G}}^2}}\left[(D_iW){K^{ij}}^{-1}(D_jW)^*-\frac{3}{M^2_G}|W|^2 \right] \, , \label{sugra}
\end{equation}
where $K$ is the K\"ahler potential and $W$ is the superpotential. 
Here we have defined 
\begin{eqnarray}
K^{ij} &=& \frac{\partial^2K}{\partial{\phi}_i \partial{\phi}^*_j} \, ,  \\
D_iW   &=& \frac{\partial W  }{\partial{\phi}_i  }+\frac{1}{M^2_G}\frac{\partial K}{\partial{\phi}_i}W \, , 
\end{eqnarray}
and $M_{G}=1/\sqrt{8\pi G}$ is the reduced Planck mass.
During the standard slow-roll inflation, the energy density $\rho$ is
dominated
 by the potential energy, so the Friedmann equation reads
 $3 H^2 M_G^2=\rho \cong V$. 
For $K \ll M_G^2$ we find
\begin{equation}
3M_{G}^2H^2 \cong V \cong (D_iW){K^{ij}}^{-1}(D_jW)^*-\frac{3}{M^2_G}|W|^2  
\end{equation}
in F-term inflation.
Since the kinetic terms of the scalar fields are given by 
$-K^{ij}\partial_\mu \phi_i \partial^\mu \phi_j^*$,
the minimal model assumes $K=|\phi_i|^2$ which yields canonical kinetic terms. 
Here we can expand the exponential factor after canonical normalization
and diagonalization, 
\begin{equation}
e^{\frac{K}{{M_{\rm G}}^2}}=1 + \frac{\phi_i \phi^{i*}}{M_{\rm G}^2}+\cdots. 
\end{equation}
Therefore, the scalar potential $V$ contains the term $3H^2 \phi_i \phi^{i*}$. 
This means that the field $\phi_i$ receives an additional Hubble-scale mass, which
is called the Hubble-induced mass\footnote{
The inflaton should not acquire the Hubble induced mass; 
otherwise the slow-roll condition 
cannot be satisfied. 
In order to realize inflation, one must assume shift symmetry \cite{Kawasaki:2000yn} 
or adopt contrived models \cite{Murayama:1993xu}. }. 
If the Hubble parameter  is large enough at the beginning of inflation, 
these fields fall down to the origin quickly and the symmetry 
is restored during inflation. 
Moreover, in inflation models in which the Hubble parameter changes
gradually during inflation, such as the 
chaotic inflation model \cite{Linde:1983gd}, 
it is possible for the phase transition to take place during inflation.
Here we adopt the following superpotential,
\begin{eqnarray}
W&=&W_I+W_S,  \\
W_I &=& M \phi X,  \\
W_S&=&m^2s-\lambda s\psi \bar{\psi}, 
\end{eqnarray}
and the K$\ddot{\mathrm{a}}$hler potential
\begin{equation}
K=\frac{1}{2}(\phi+\phi^*)^2 +|s|^2+|X|^2+|\psi|^2+|\bar{\psi}|^2 \, , 
\end{equation}
to realize chaotic inflation, 
where $W_I$ and $W_S$ are superpotentials of the inflation 
sector and the string sector, respectively.  
Here $\phi$ is the inflaton, $X$ is an additional singlet,  $\psi$ 
and  ${\bar \psi}$ are the 
symmetry-breaking Higgs fields that break $U(1)_S$ symmetry, 
$s$ is a singlet which destabilizes the Higgs field at the origin, 
$M$ is the inflaton mass, $m$ is the symmetry breaking scale, and 
$\lambda$ is a coupling constant. 
Note that we impose $R$-symmetry and discrete $Z_2$ symmetry in order to suppress all other unwanted 
interactions such as $X\phi^2, s \phi$, and so on\footnote{Note 
that the term $\epsilon s \phi^2$ in the superpotential cannot be forbidden by these symmetries 
but it breaks the shift symmetry of $\phi$. 
Since we can expect that  the inflaton mass $M$ arises as an order parameter of the shift symmetry breaking,
the coupling constant $\epsilon$ should be suppressed enough, say $\epsilon\sim M^2\sim 10^{-10}$
in Planck units. Thus, the following discussion does not change. }. 
Charge assignments on the fields are shown in Table~\ref{tab1}. 
\begin{table}[t!]
  \begin{center}
    \begin{tabular}{ | c | c | c | c | c |c|}
      \hline 
         ~          &  $\phi$  & $X$ & $s$ & $\psi$ & ${\bar \psi}$\\ \hline
        $U(1)_S$ & 0 & $0$ & $0$ &  $+1$ & $-1$ \\ \hline
        $R$ & 0 & $+2$ & $+2$ & 0 & 0 \\ \hline
        $Z_2$  & $-1$ & $-1$ & 0 & 0 & 0 \\ \hline   
    \end{tabular}
    \caption{ 		
     	Charge assignments on superfields in the model under the $U(1)_S$ symmetry, 
	the $R$-symmetry, and $Z_2$-symmetry. \label{tab1}
     }
  \end{center}
  \label{table:charge}
\end{table}

This model can realize chaotic inflation naturally 
thanks to the shift symmetry, $\phi \rightarrow \phi+c$ with $c$ being 
a real parameter,  along the imaginary part of $\phi$
which plays the role of the inflaton \cite{Kawasaki:2000yn}. 
Setting $\phi=\frac{1}{\sqrt{2}}(\chi+i\varphi)$, 
the potential for these fields is
\begin{eqnarray}
V &=& e^{\frac{K}{{M_{\rm G}}^2}}\left[(D_iW){K^{ij}}^{-1}(D_jW)^*-\frac{3}{M^2_G}|W|^2 \right] \nonumber \\
&=& \exp \left[\frac{1}{M_{\rm G}^2}\left( \chi^2+ |X|^2 +|s|^2 +|\psi|^2 +|\bar{\psi}|^2 \right) \right]  \nonumber \\
&\:& \times \left[ \frac{1}{2}M^2(\chi^2+\varphi^2) +M^2|X|^2 +m^4 -\lambda m^2(\psi \bar{\psi}+ \psi^* \bar{\psi}^*) \right. \nonumber \\
&\;& \qquad +\lambda^2 |\psi|^2 |\bar{\psi}|^2 +\lambda^2 |s|^2 |\bar{\psi}|^2 +\lambda^2 |s|^2 |\psi|^2 \nonumber \\
&\;& \; \; \; \; +\frac{1}{M_{\rm G}^2} \left. \left\{  \left( \frac{3}{2}M^2\chi^2-\frac{1}{2}M^2\varphi^2 \right)|X|^2+\cdots \right\} \right] \,.
\end{eqnarray}
Note that we do not have to consider all of these terms because 
the fields $\chi$, $s$, $X$, $\psi$, and $\bar{\psi}$ have very steep potential
due to the factor $e^{\frac{K}{{M_{\rm G}}^2}}$ at the field values larger than $M_G$.
This means they cannot have field values much bigger than $M_{\rm G}$.
Since $\chi$ and $s$ have larger masses than the Hubble
parameter, they are 
practically fixed at the origin during inflation.
Although $X$ has a mass of the order of the inflaton mass, 
its contribution to the total potential energy (or the dynamics of inflation) 
and the density fluctuation are negligible 
due to its much smaller field value than the inflaton.
Since only the inflaton $\varphi$ can have a value much larger than $M_{G}$, 
this model reduces to a simple chaotic inflation model, $V=M^2 \varphi^2/2$. 

Now we investigate the phase transition during inflation. 
Diagonizing the mass matrix of $\psi$ and $\bar{\psi}$, we obtain the following canonically normalized fields:
\begin{eqnarray}
  \left\{
    \begin{array}{l}
      \psi_1=\dfrac{1}{\sqrt2}(\psi+{\overline{\psi}}^*) \, , \\
      \psi_2=\dfrac{1}{\sqrt2}(\psi-{\overline{\psi}}^*) \, .
    \end{array}
  \right.
\end{eqnarray}
The relevant terms that control the dynamics of the phase transition during
inflation are
\begin{eqnarray}
V &=& \frac{1}{2}M^2 \varphi^2 + 
\left[ \frac{1}{M^2_G} \Bigl( \frac{1}{2}M^2{\varphi}^2+m^4 \Bigr)-\lambda m^2\right]|\psi_1|^2 +\frac{\lambda^2}{4}|\psi_1|^4  \nonumber \\
& \approx &\frac{1}{2}M^2 \varphi^2 + 
\left(3H^2 -\lambda m^2 \right)|\psi_1|^2 +\frac{\lambda^2}{4}|\psi_1|^4 \,.
\end{eqnarray}
The first term is the potential for the inflaton. 
The mass term of $\psi_1$ changes its sign from positive to negative during inflation
at $H^2=\lambda m^2/3$. 
As a result, $\psi_1$ may be destabilized and the $U(1)_S$ symmetry is 
spontaneously broken during inflation, while $\psi_2$ has always
positive mass squared.

Our model contains the parameters $\lambda$, $m$, and $M$. 
$M$  is related to the primordial power spectrum of curvature fluctuations via 
\begin{equation}
\Delta_{\cal R}^2(k)=\left. \frac{1}{24\pi^2}\frac{V}{M_{\rm G}^4}\frac{1}{\epsilon}\right|_{k=aH}
\approx \frac{1}{6\pi^2}\frac{M^2}{M_{\rm G}^2}{\cal N}^2, 
\end{equation}
with ${\cal N}\simeq 55$ being the number of $e$-folds when observable scales exit the horizon. 
The observed value $\Delta_{\cal R}^2(k)=2.4\times 10^{-9}$\cite{Komatsu:2010fb} 
determines the inflaton mass to be 
$M \cong 10^{13}$GeV.
We assume $\lambda$ is of order of unity and $m$ is much smaller than $M_{\rm G}$.

The line energy density of cosmic strings is related to 
the vacuum expectation value of the string-generating field as
\footnote{To be precise, one should treat the coefficient as
a function depending on $\lambda$, rather than as a constant, $2\pi$.
However within the range of interesting values of $\lambda$ in our model,
we can approximate this coefficient as 2$\pi$\cite{Jeannerot:2005mc}.
}
\begin{equation}
\mu \cong 
2\pi \langle {\psi_1}^2 \rangle = 4\pi\frac{m^2}{\lambda} \, .
\end{equation}
Therefore one can determine the value of $m^2/\lambda$ by fixing $G\mu$.

The crucial issue is whether strings with $G\mu \geq 10^{-8}$, whose
tension is large enough to be detectable by CMB observations 
\cite{Foreman:2011uj}, can be
produced with appropriate density so that they can evade the PTA
constraint but are observable using CMB.  
It is therefore important to study when 
the phase transition takes place to clarify the mean separation and
the correlation length of the string network.
 
In our scenario, the phase transition is triggered by quantum
fluctuation
during inflation rather than thermal fluctuations and we can make use
of the results of \cite{Yokoyama:1989pa,VOS,Nagasawa:1991zr,
Kamada:2011bt} to analyze its properties.   
Calculations based on the stochastic inlfation 
method \cite{Starobinsky:1986fx,Starobinsky:1994bd}
show that the phase of
the string-forming field $\psi_1$ is fixed when the classical potential
force surpasses the effect of stochastic quantum fluctuations
\cite{Nagasawa:1991zr}.  This occurs $\Delta{\cal N}=\sqrt{c/2}$ e-folds
after $\psi_1=0$ has become classically unstable
with $3H^2=\lambda m^2$ where $c\equiv \lambda m^2/M^2$ 
\cite{Nagasawa:1991zr,Kamada:2011bt}.
The number of e-folds from this epoch to the end of inflation is given
by
\begin{equation}
{\cal N}_D= \frac{1}{2} (c-\sqrt{2c}-1) \,, 
\end{equation}
which gives the extra dilution which is absent in the case where phase transitions
occur after or without inflation.
The Hubble parameter at this epoch, which we denote with a suffix $f$,
is given by 
\begin{equation}
H_f=\left(\frac{c}{3}-\frac{\sqrt{2c}}{3}\right)^{1/2}M.
\end{equation}

The typical separation between strings at this epoch
can be estimated as in 
\cite{Yokoyama:1989pa,VOS}.
First, choose an arbitrary point $\mbox{\boldmath $x$}_0$ with an arbitrary field value of
$\psi_1(\mbox{\boldmath $x$}_0)$.
One can always perform a gauge transformation to make this real: 
$\psi_1(\mbox{\boldmath $x$}_0)\equiv \psi_{1R}(\mbox{\boldmath $x$}_0)+i\psi_{1I}(\mbox{\boldmath $x$}_0)=\psi_{1R}(\mbox{\boldmath $x$}_0)$.
Then, since a string is
a locus of $\psi_{1R}(\mbox{\boldmath $x$})=\psi_{1I}(\mbox{\boldmath $x$})=0$, the nearby string segment can be
found by searching for a point (in fact, a line) with $\psi_{1R}(\mbox{\boldmath $x$})=0$ on the
surface of $\psi_{1I}(\mbox{\boldmath $x$})=0$.  To the lowest-order Taylor expansion, the
condition for a string to exist within a distance $r$ from the point
 $\mbox{\boldmath $x$}_0$ is given by
\begin{equation}
  r |{\vect e}_I\times\nabla\psi_{1R}(\mbox{\boldmath $x$}_0)|>|\psi_{1R}(\mbox{\boldmath $x$}_0)|,
\end{equation}
where $\vect e_I \equiv \nabla\psi_{1I}(\mbox{\boldmath $x$}_0)/|\nabla\psi_{1I}(\mbox{\boldmath $x$}_0)|$
is the unit normal vector to the surface $\psi_{1I}=0$.
Since $\psi_{1R}(\mbox{\boldmath $x$}_0)$ and $\nabla\psi_{1R}(\mbox{\boldmath $x$}_0)$ are statistically independent,
almost Gaussian variables, one can express the probability
for the inequality to be satisfied at a distance $r$, $P(r)$,
in terms of the variances $\langle\psi_{1R}^2\rangle\equiv \sigma^2$
and $\langle (\nabla\psi_{1I})^2\rangle/3
=\langle (\partial_x\psi_{1I})^2\rangle\equiv \sigma^2_{g'}$, which 
will be evaluated later.  
Without loss of generality one can take
$z$-axis along $\vect e_I$.  Then writing $\nabla\psi_{1R}\equiv
(u_x,u_y,u_z)$, $P(r)$ is given by
\begin{align}
 P(r)&=\int_{r^2(u_x^2+u_y^2)>\psi_{1R}^2}P(\psi_{1R},u_x,u_y,u_z)d\psi_{1R}d^3u
\nonumber \\
&=\frac{1}{(2\pi)^2\sigma\sigma_{g'}^3}\int\exp\left(
-\frac{\psi_{1R}^2}{2\sigma^2}-\frac{u_x^2+u_y^2+u_z^2}{2\sigma_{g'}^2}\right)
d\psi_{1R}d^3u \\
&=\frac{\sigma_{g'}r}{(\sigma^2+\sigma_{g'}^2r^2)^{1/2}}, \nonumber
\end{align}
where $P(\psi_{1R},u_x,u_y,u_z)$ is the one-point probability distribution
function
of $\psi_{1R}$ and $\nabla\psi_{1R}$.
Thus the typical distance to reach a string, which also gives an
estimate of the mean separation of strings,
 is given by $r\simeq \sigma/\sigma_{g'}$. 

Next we calculate the variances at $t=t_f$, which is obtained from
the power spectrum of $\psi_1$  \cite{Kamada:2011bt}
\begin{equation}
{\cal P}_{\psi_1} = |\psi_{1k}(t_f)|^2 \simeq \frac{H^2(t_k)}{2k^3} \left(\frac{S(t_k)}{S(t_f)}\right)\exp \left[-\frac{9}{2M^2}\left(H(t_k)-\frac{2 M}{3}\sqrt{c-\frac{1}{2}}\right)^2+\frac{(2-\sqrt{3})^2c}{4}\right]\,, 
\end{equation}
where
\begin{equation}
S(t) \equiv \left( \lambda m^2-\frac{M^2}{2}-\frac{3H^2(t)}{4}\right)^{1/2}\,,
\end{equation}
and $t_k$ is the epoch when the comoving mode $k$ exits the horizon.
We can evaluate the ratio of the variances as
\begin{align}
\label{r^2}
\frac{\sigma^2}{\sigma^2_{g'}}
&= \frac{\int|\psi_{1k}(t_f)|^2\frac{d^3k}{(2\pi)^3}}{\int \frac{k^2}{3}|\psi_{1k}(t_f)|^2\frac{d^3k}{(2\pi)^3}} 
\equiv \left( \frac{\tilde{r}(c)}{H_f} \right)^2 \nonumber\\
&\simeq 15\sqrt{2\pi} (x_f-x_d)\exp\left[\frac{15}{2}(x_f-x_d)^2+\frac{c}{5}-\sqrt{2c}+\frac{2}{5}  \right]\frac{1}{H_f^2}
\end{align}
where 
\begin{equation}
x_d\equiv \frac{2}{5}\left(c-\frac{1}{2}\right)^{1/2},~~x_f\equiv\left(\frac{c}{3}-\frac{\sqrt{2c}}{3}\right)^{1/2}.
\end{equation}
Now we get the typical separation of strings.
In addition to the expression (\ref{r^2}), we also numerically evaluate $\tilde{r}(c)$ as a function of $c$
and get $\tilde{r}(c)\simeq 400$ at $c=100$.

After the initial configuration of the string network is determined at $t_f$ during inflation,
the typical separation simply scales as the scale factor until it falls below the Hubble radius.
At redshift $z$, it reads
\begin{eqnarray}
\label{d at z}
d(z)&=& r e^{{\cal N}_D}\frac{a_{\rm R}}{a_{\rm end}}\frac{a_{\rm eq}}{a_{\rm R}}\frac{a(z)}{a_{\rm eq}} \nonumber \\
&=& 3\times 10^{18}\frac{\tilde{r}(c)}{x_f(c)}\exp\left[\frac{1}{2}(c-\sqrt{2c }-1 )\right] 
\frac{1}{z+1}\,\left( \frac{T_{\rm R}}{10^6 {\rm GeV}}\right)^{-\frac{1}{3}} {\rm GeV}^{-1} ,
\end{eqnarray}
where the suffixes end, R, and eq represent the epochs of the end of inflation, 
reheating and radiation-matter equality, respectively.
If we assume the network enters the scaling
regime immediately after the string separation becomes smaller than the characteristic length
of a scaling network $d(z)\lesssim \xi(z)$, where,
\begin{equation}
\label{gammadH}
\xi(z) =\gamma d_{\rm H}=\gamma H^{-1}=\frac{\gamma \times 7\times
10^{41}}{\sqrt{\Omega_{\rm \Lambda 0}+\Omega_{\rm m
0}(1+z)^3+\Omega_{\rm r 0}(1+z)^4}}{\rm GeV}^{-1} ,
\end{equation}
then we can get a relation between the model parameres and $z_{\rm sc}$ by equating (\ref{d at z}) and
(\ref{gammadH}) to determine the region of the parameter space where traces of cosmic strings are
detectable by future CMB observations without conflicting with the PTA constraints.
Figure 3 shows allowed region of the model parameters where the redshift at the onset
of the scaling regime has been obtained from an equality
\begin{equation}
d(z_{\rm sc})/\xi(z_{\rm sc})=1.
\end{equation}

We must admit, however, that this estimate has some uncertainty. 
Unlike in the
standard scenario where the string network approaches the scaling solution as the
string density decreases, in the present case it is
reached as the number of long strings per the Hubble volume increases to the scaling value. 
To our knowledge, such a situation has not been numerically simulated,
and it is an open issue when the system relaxes to the scaling solution. 
Fortunately, however, there exists a finite section of parameter space where all the requirements are met, 
even if we change our criterion of the epoch when the scaling law is achieved.

Indeed, if we assume the scaling is achieved when an equality $d(z_{\rm sc})/\xi(z_{\rm sc})=K$
is satisfied, we find that the two vertical lines and the oblique line representing the
gravitational-wave constraint in Figure 3 shift to the right by
\begin{align}
&\frac{2}{100}\left(\frac{\sqrt{2c}}{\sqrt{2c}-1} \frac{c-\sqrt{2c}}{c-\sqrt{2c}-1} \right)\log K\\
\simeq&\, 2.2\times10^{-2}\log K \quad({\rm at}\, c=100) \notag
\end{align}
so that the shape and the size of the allowed region remain intact. 
Note that the parameter
$K$ can also accommodate possible variation of the initial string separation, too.
Therefore, we conclude that there are certainly interestingly large parameter space
where signatures of the cosmic strings are detectable by the future CMB observation
evading the PTA constraints.

We have shown in Figure 3 the parameter corresponding to 
$z_{\rm sc}=1100$ and $d(z=0)=H_0^{-1}$ which means
there would be only one string inside our present horizon.
We find $m$ should be about $10^{14}$GeV, and so use $m_{\rm 14} \equiv m/10^{14}$GeV.
Therefore $c$ becomes $\displaystyle{100\lambda m_{14}^2}$.
The horizontal axis is related to the time when strings start to scale and produce loops.
According to the results of the previous section, an upper limit is imposed on $z_{\rm sc}$
in order not to contradict the MSP observation.
This upper limit on $z_{\rm sc}$ yields a lower limit on $\lambda m_{14}^2$.
The value of $G\mu$ is proportional to the value of vertical axis.

\begin{figure}[h]
 \begin{center}
  \includegraphics[width=150mm]{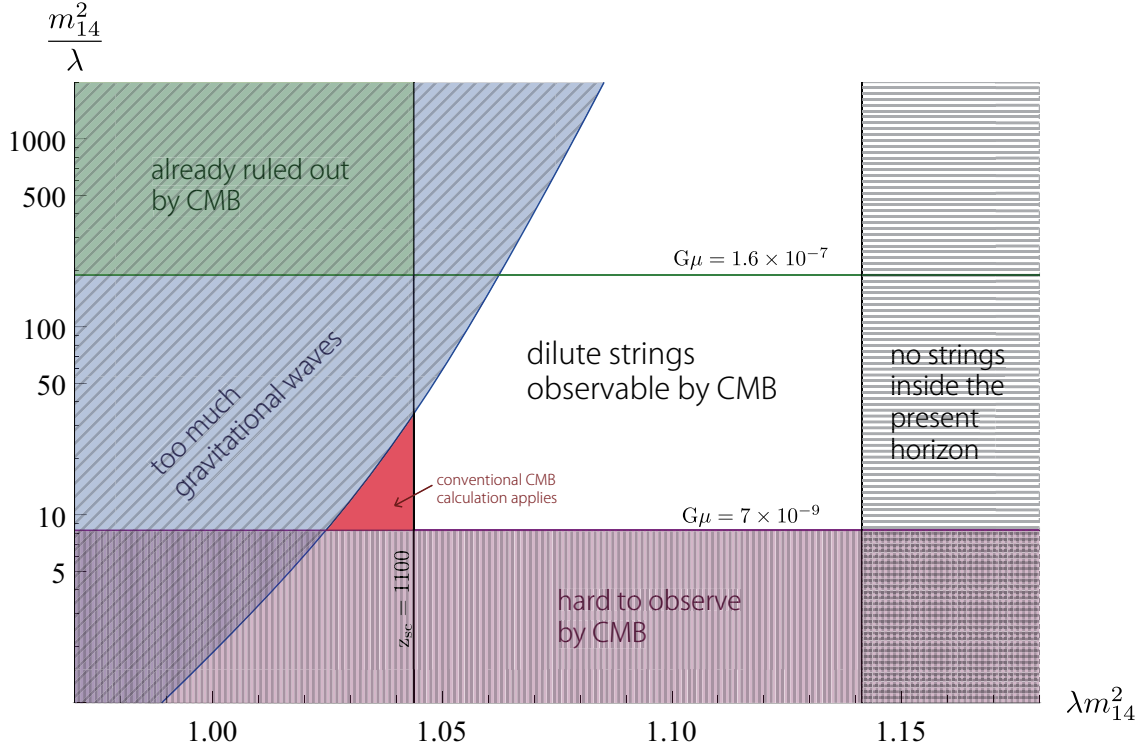}
 \end{center}
 \begin{minipage}[h]{0.95\columnwidth}
 \caption{The allowed parameter region of this model is shown. 
 In the blue-painted region, the network of strings starts to produce loops so early 
 that the gravitational waves emitted by them contradict observations.
 Though determining this region requires more statistical analysis of bursts and a more realistic 
 treatment of $\alpha$, we emphasize that some region remains allowed.
 The most interesting region, which represents the case
 strings in which can not only evade the PTA constraint 
 but also contribute to the anisotropy of the CMB as much as ordinary strings do.
 The white region corresponds to the case where strings are observable through the CMB but at densities 
 smaller than in the conventional scenario.
 As a recent constraint on $G\mu$, we used the value obtained in \cite{Dunkley:2010ge}.
 According to \cite{Foreman:2011uj}, strings with $G\mu<7\times 10^{-9}$ are not observable.
 }
 \end{minipage}
\label{fig:region}
\end{figure}

\section{Discussion \label{sec5}}
In this paper, we have studied a new mechanism to evade the PTA constraints on the cosmic string tension. 
We have found that the PTA constraints can be evaded if the loop formation starts at a later epoch, say,  
$z\lesssim 10^{4}$, depending on the value of $\alpha$ and the tension of the cosmic strings. 
We have also shown that this scenario can be realized if the phase 
transition and string formation 
take place during inflation.  
As a specific example, we have shown there is a finite
region of parameter space 
where we can expect to detect the signatures of cosmic strings through CMB observations 
while accommodating the PTA constraints in the F-term
chaotic inflation model in supergravity \cite{Kawasaki:2000yn}.

As seen in Figure 1, the value of $\alpha$ is very important to
determine the constraint on $z_{\rm sc}$.
Recent simulation \cite{BlancoPillado:2011dq} indicates that 
loops are formed with various sizes ranging from $\alpha \simeq
10^{-1}$ to $\alpha < 10^{-5}$ in the standard scenario.
In our scenario, however, the mean separation of cosmic strings 
is much larger than the horizon initially, 
so it is expected that larger loops are produced more frequently 
than in the ordinary scenario.
In addition, inflation also erases small-scale fluctuations on long strings. 
Hence  strings do not develop small-scale structure 
and the correlation length remains of the order of the Hubble radius for some time, 
even after $z_{\rm sc}$. As a result, we expect $\alpha$ to take 
a relatively large value, $\alpha \sim 10^{-1}$, at the onset of the 
scaling regime.

Though we assume that the cosmic string network enters the scaling regime
when the mean separation of cosmic strings becomes comparable to the
characteristic length of the scaling network, the time of the onset of
the scaling regime is still ambiguous. 
However, we can expect that the earliest candidate for the
onset of the scaling regime is when the smallest scale of the cosmic strings,
$(a(z)/a_f)H_f^{-1}$, 
becomes comparable to the characteristc length of the scaling network $\xi(z)$.  
We confirmed that, even in this case, there is a
finite but interesting parameter space. 
Therefore, we conclude that there is certainly a region of
parameter space where we can expect to detect the signatures of cosmic strings
by future CMB observations.

In our scenario, the effect of the CMB may need to be reconsidered. 
If inflation did not last so long after strings were formed, 
stings start to satisfy the scaling rule at an earlier epoch. 
As a consequence,
their separation would already become
 smaller than the horizon at the recombination.
Then there would be as many strings as in the standard scenario 
near the last scattering surface of the CMB photons.
The effect on the CMB would be the same as that has already been 
analyzed in the literature 
\cite{Hindmarsh:1993pu}.
On the other hand,  if inflation lasted 
very long after strings had been formed, 
they could not start to scale until very late time. 
Therefore, their separation would be larger than the 
horizon at the recombination. 
In this case, their effect on the CMB would become smaller.  
For a quantitative understanding, numerical simulations of
the string network with appropriate
initial conditions are desired.

\section*{Acknowledgments}
We would like to thank Christophe Ringeval and Matthew Lake for helpful comments.
This work was partially supported by JSPS Grant-in-Aid for Scientific
Research No. 23340058 (J.Y.), Grant-in-Aid for Scientific
Research on Innovative Areas No. 21111006 (J.Y.),
and Global COE Program ``the Physical
Sciences Frontier", MEXT, Japan.


\end{document}